\documentclass[10pt,twocolumn,english,aps,twocolumn,showpacs,superscriptaddress,XXlinenumbers]{revtex4-1}
\usepackage[T1]{fontenc}
\usepackage[latin9]{inputenc}
\usepackage{color}
\usepackage{babel}
\usepackage{amsmath}
\usepackage{amssymb}
\usepackage{graphicx}
\usepackage[unicode=true,pdfusetitle,
 bookmarks=true,bookmarksnumbered=false,bookmarksopen=false,
 breaklinks=false,pdfborder={0 0 0},pdfborderstyle={},backref=false,colorlinks=true]
 {hyperref}
\hypersetup{
 citecolor=blue}

\makeatletter

\newcommand{\noun}[1]{\textsc{#1}}

\graphicspath{{./}{./FIGURES/}} 
\usepackage{babel}

\makeatother

\begin{document}

\title{Nonlocality Induces Chains of Nested Localized Structures}

\author{J. Javaloyes}

\affiliation{Departament de Física, Universitat de les Illes Baleares, C/Valldemossa
km 7.5, 07122 Mallorca, Spain}

\author{M. Marconi}

\affiliation{Université Côte d'Azur, CNRS, Institut de Physique de Nice, F-06560
Valbonne, France}

\author{M. Giudici}

\affiliation{Université Côte d'Azur, CNRS, Institut de Physique de Nice, F-06560
Valbonne, France}
\begin{abstract}
Localized Structures often behave as quasi-particles and they may
form molecules characterized by well-defined bond distances. In this
paper we show that pointwise nonlocality may lead to a new kind of
molecule where bonds are not rigid. The elements of this molecule
can shift mutually one with respect to the others while remaining
linked together, in a way similar to interlaced rings in a chain.
We report experimental observations of these chains of nested localized
structures in a time-delayed laser system.

\end{abstract}
\maketitle

In a seminal contribution, Alan Turing set the bases of morphogenesis
\cite{turing52}. He demonstrated that, in a dissipative environment,
the interplay between \emph{local} nonlinearities and differential
operators was sufficient to initiate self-organization and generate
an infinite variety of patterns. These emergent structures can be
found in many physical, biological, and laboratory systems. Among
them, localized structures (LSs) are of particular interest and have
been widely observed in nature \cite{WKR-PRL-84,MFS-PRA-87,NAD-PSS-92,UMS-NAT-96,AP-PLA-01}.
These states can be individually addressed by a local perturbation,
without affecting their surrounding environment. They are particularly
relevant for applications when implemented in optical resonators as
light bits for information processing \cite{RK-OS-88,FS-PRL-96,BLP-PRL-97,1172836,BTB-NAT-02,LCK-NAP-10,GBG-PRL-08,TAF-PRL-08,HBJ-NAP-14}.
Localized Structures may form bound states, also called ``molecules'',
via the overlap of their oscillating tails which creates ``covalent''
bonds corresponding to stable equilibrium distances \cite{RK-JOSAB-90,AP-PLA-01,MFH-PRE-02}.

In this letter we disclose a different kind of molecule composed by
chains of nested LSs, which are globally bounded yet locally independent,
and similar to an ensemble of interlaced rings. Interesting enough,
similar molecular structures exist in chemistry. They are composed
by interlocked macrocycles and they are called catenanes \cite{FW-JAC-61}.
While usual covalent bound states of LSs move as a rigid ensemble
when subject to perturbations, the stability analysis of these interlaced
LSs reveals that they exhibit several neutral modes corresponding
to the individual translation of each element. In other words, small
displacements between the components do not relax, yet the ensemble
remains stable. We show that these molecules, which challenge the
usual notion of local stability for LS compounds, can be obtained
in presence of a pointwise nonlocality coupling a field $\Phi\left(x,t\right)$
to a distant point in space $\Phi\left(x+a,t\right)$.

Nonlocality has been widely explored in spatially extended systems
and has been shown to induce patterns \cite{RDA-PRL-98}, convective
instabilities \cite{PZ-PRL-05,ZP-PRL-07} or N-fold structures \cite{RNP-EPL-03}.
Distributed nonlocality was recently found capable of stabilizing
LSs \cite{FCE-PRL-13} and identified as an important mechanism governing
the morphogenesis processes in liquid crystal \cite{TFC-PRA-15} and
vegetation patterns \cite{EFC-PRE-15}. Global coupling, as an extreme
case of nonlocality, is known to have a deep impact on LSs bifurcation
diagrams \cite{FCS-PRL-07}.

\begin{figure}[b]
\centering{}\includegraphics[bb=0bp 0bp 7491bp 2491bp,clip,width=1\columnwidth]{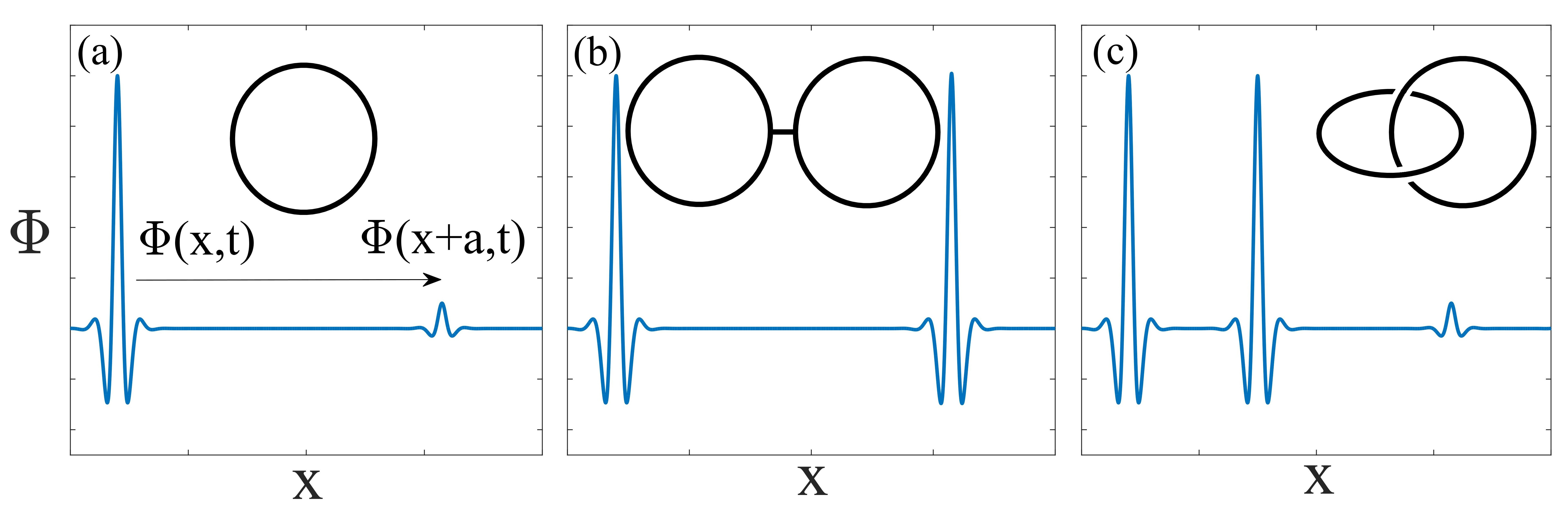}\caption{(a) Sketch of a LS with a pointwise nonlocal term where the field
$\Phi\left(x,t\right)$ induces a perturbation at a distance $a$.
(b) Covalent molecule where bonding occurs via tail interactions.
(c) Molecule with a nested element.\label{fig:sketch}}
\end{figure}

The difference between standard covalent molecules and the nested
states is illustrated in the qualitative sketches of Fig.~\ref{fig:sketch}.
We represent in Fig.~\ref{fig:sketch}a) an elementary LS solution
stemming from a generic partial differential equation (PDE) able to
sustain LSs in one spatial dimension $\left(x\right)$. The characteristic
length of the LS is denoted $l$. A small pointwise nonlocal perturbation
can be modeled for instance as a linear perturbation of a PDE that
reads
\begin{eqnarray}
\partial_{t}\Phi & = & \mathcal{F}\left(\left|\Phi\right|^{2},\partial_{x}^{2}\right)\Phi+\varepsilon\Phi\left(x+a,t\right)
\end{eqnarray}
 with $\mathcal{F}$ an operator representing e.g. the Ginzburg-Landau
equation as in \cite{TF-JPF-88,FT-PRL-90}. Without loss of generality
we assume $a>0$. For sufficiently small $\varepsilon,$ the overall
structure of the original LS remains preserved, but it develops a
small echo of amplitude $\varepsilon$ at the distance $a$, as a
consequence of the nonlocal term. Additional weaker replica of this
echo also appear at distances $na$ ($n\in\mathbb{N})$ with amplitudes
$\varepsilon^{n}$, but they can be neglected in this discussion.
If $\varepsilon=0,$ standard covalent molecules can be generated
by the interaction between the decaying oscillating tails of two nearby
LSs. This leads to equilibrium distances $d\sim l$ (not shown). When
$\varepsilon\neq0$, and for moderate nonlocality $a\sim l$, the
rightmost LS tail will be modified, which may lead to additional equilibrium
distances that persist even when $a\gg l$. Here the binding occurs
as a consequence of the interaction between the main pulse of a second
LS and the echo of the first, leading to an equilibrium distance $d\sim a$
, see Fig.~\ref{fig:sketch}b). All these rigid molecules feature
well-determined bond lengths and a single neutral mode corresponding
to the translation degree of freedom of the whole ensemble. A novel
kind of molecule appears in the case $a\gg l$, when the second LS
is placed at a distance $d$ from the first LS such that $l<d<a$.
In this case, shown in Fig.~\ref{fig:sketch}c), the two LSs are
sufficiently far so that they can move independently but, at the same
time, the second LS cannot overcome the repulsive barrier induced
by the echo of the first LS. As such, the two LSs are globally linked
while being locally independent. A useful analogy to picture these
situations can be made using rings which are rigidly bounded in the
situation represented in Fig.~\ref{fig:sketch}b), while they are
interlaced in the case of Fig.~\ref{fig:sketch}c).

Pointwise nonlocality may not be easy to achieve experimentally and,
in order to observe such new bounded states, we have studied their
realization in a time-delayed system (TDS). In recent years, building
on the strong analogies between spatially extended and time-delayed
systems \cite{AGL-PRA-92,GP-PRL-96,K-CMMP-98}, the latter have been
proposed for controlling spatial LSs \cite{G-PRE-13}, hosting chimera
states \cite{LPM-PRL-13}, domain walls \cite{GMZ-EPL-12,GMZ-PRE-13,MGB-PRL-14,JAH-PRL-15},
vortices \cite{YG-PRL-14} and, in particular, temporal LSs \cite{MJB-PRL-14,GJT-NC-15,RAF-SR-16,MJB-NAP-15,J-PRL-16},
see \cite{YG-JPA-17} for a review. The idea that a time delay $\tau$
is akin to a spatial dimension is rooted in the representation developed
in \cite{AGL-PRA-92} that consists in cutting a single temporal time
trace generated by a TDS into $n\in\mathbb{N}$ chunks of duration
$T$ (with $T\simeq\tau$) and stack them into a two dimensional map.
The horizontal dimension plays the role of a pseudo-spatial variable
representing the profile of the pulse within the n-th period while
the vertical axis depicts the discrete time index $n$. In some situations
\cite{GP-PRL-96}, this two-dimensional representation can even lead
to an analytical description of a TDS as a PDE containing \emph{local}
operators, as e.g. Laplacian. In this formalism, the temporal profile
over one period \textendash the information in speudo-space\textendash{}
is mapped onto the next period. As such, one understands that the
inclusion of a second delay $\tau_{2}=a$ induces a pointwise nonlocality
coupling each point of the temporal profile to itself, but with a
shift $a$. In addition, similar results are to be found for $\tau_{2}=pT+a$
if $p\in\mathbb{N}$ is sufficiently small so that the temporal profile
does not evolve much during $p$ periods.

\begin{figure}[t]
\begin{centering}
\includegraphics[clip,width=1\columnwidth]{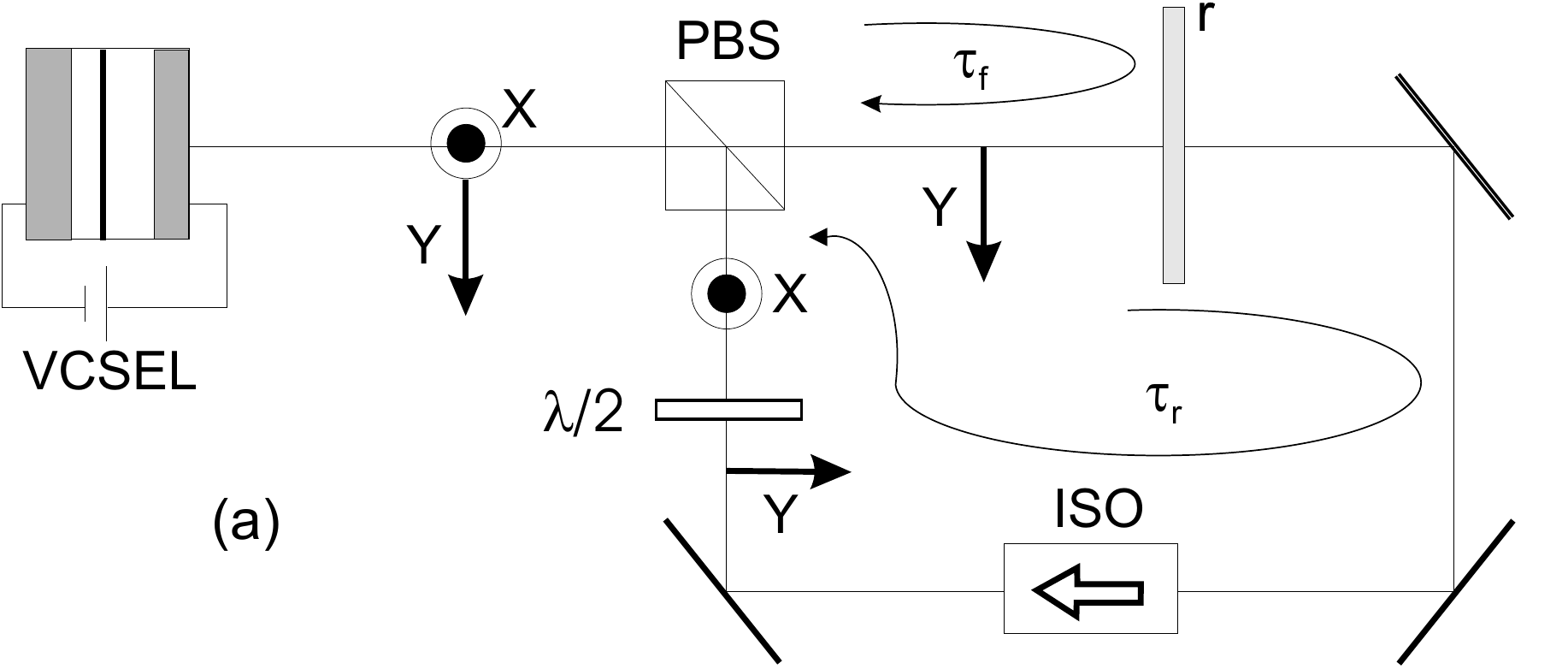}
\par\end{centering}
\centering{}\includegraphics[bb=20bp 0bp 400bp 100bp,clip,width=1\columnwidth]{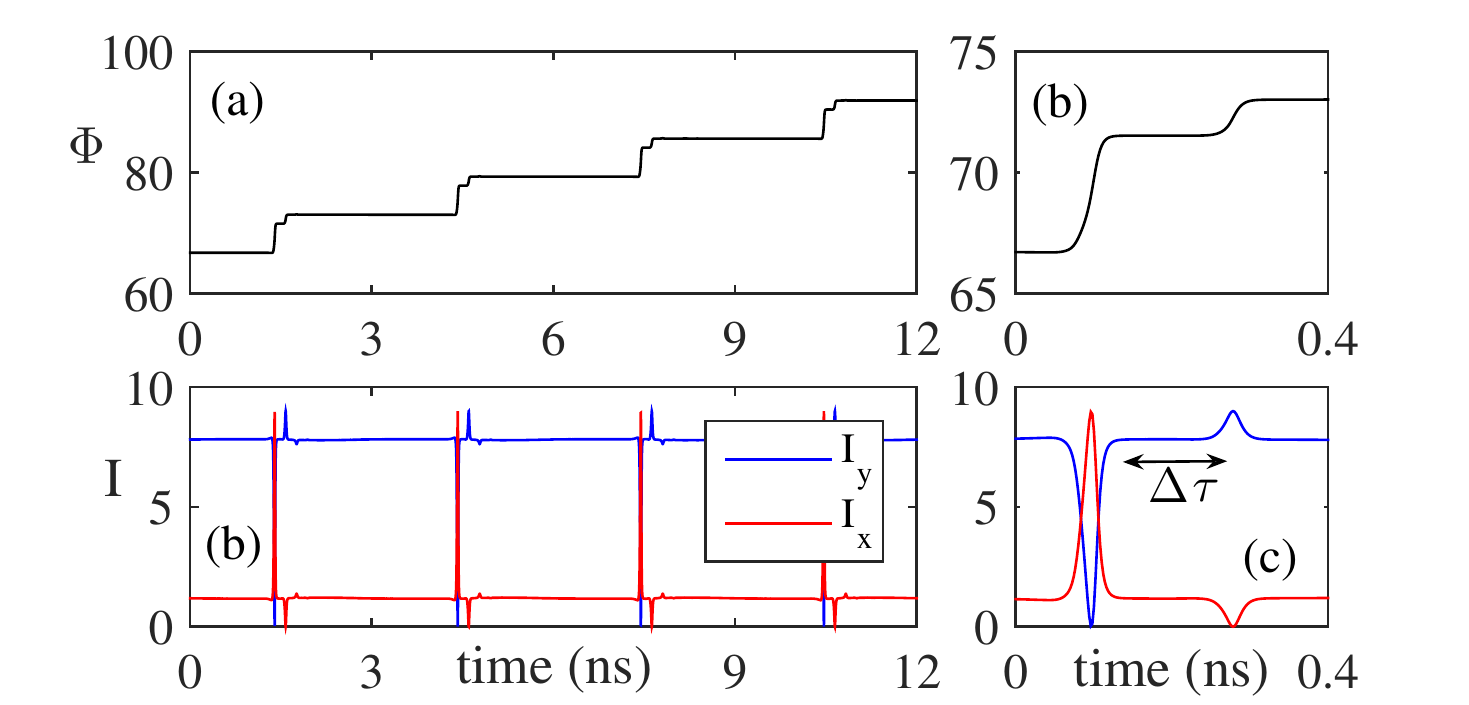}\caption{(a) Experimental setup. The polarizing beam splitter (PBS) splits
the X and Y components of the field. $r$ is a partially reflective
mirror that feeds back the Y component after a time of flight $\tau_{f}$.
The optical isolator (ISO) only allows for the transmission of the
field traveling from right to left. The half waveplate ($\lambda/2$)
rotates the transmitted Y component of the field into its orthogonal
polarization, which is then re-injected after a time delay $\tau_{r}$.
(b,c) Numerical integration of Eq.~(\ref{eq:SineDDE}). Temporal
trace corresponding to a periodic regime with a single LS (b) and
(c) close-up around the LS ($\Delta\tau=\tau_{r}-\tau_{f}$). \label{fig:setup}}
\end{figure}

An example of a photonic system with a double time delay capable of
hosting LSs and molecules has been recently described \cite{MJB-NAP-15}
and is summarized in Fig.~\ref{fig:setup}a). A single-transverse
mode vertical-cavity surface-emitting laser (VCSEL) is coupled to
an external cavity that selects one of the linearly polarized states
of the VCSEL (Y, say) and feeds it back twice, once with a time delay
$\tau_{f}$, and once with a delay $\tau_{r}$ after being rotated
into the orthogonal direction. When $\tau_{f}$ is much larger than
the laser timescales and for properly chosen parameters, this system
hosts vectorial LSs which correspond to a full rotation of the polarization
vector state on the Poincaré sphere. Along the X polarization component,
they correspond to upward pulses over a low intensity background while
they correspond to anti-phase downward pulses in the Y polarization
component, see Fig.~\ref{fig:setup}b,c). The situation depicted
in Fig.~\ref{fig:setup}a) is well described by the so-called Spin-Flip
Model \cite{SFM-PRA-95} supplemented by the inclusion of the delayed
re-injection terms. Yet, a multiple timescales analysis of this model
applied in the limit of large damping of the relaxation oscillations,
weak dichroism $\gamma_{a}$, moderate birefringence $\gamma_{p}$
and feedback rates $\left(\eta,\beta\right)$ leads to a decoupling
between the equations for the total intensity $I_{0}$, the population
inversion, the field ellipticity and the optical phase, from the equation
describing the polarization vector longitude $\Phi$ along the equatorial
plane of the Poincaré sphere \cite{JMG-PRA-14}. The following simpler
delayed equation was found to reproduce well the dynamics,
\begin{eqnarray}
\frac{\dot{\Phi}}{2} & = & z\sin\Phi+\bar{\eta}\sin\frac{\Phi^{\tau_{f}}}{2}\cos\frac{\Phi}{2}-\bar{\beta}\sin\frac{\Phi}{2}\sin\frac{\Phi^{\tau_{r}}}{2},\label{eq:SineDDE}
\end{eqnarray}
with $\Phi^{\tau_{f,r}}=\Phi\left(t-\tau_{f,r}\right)$ the delayed
arguments, $z=\alpha\gamma_{p}+\gamma_{a}$, $\left(\bar{\eta},\bar{\beta}\right)=\left(\eta,\beta\right)\sqrt{1+\alpha^{2}}$
with $\alpha$ the Henry's linewidth enhancement factor. If not otherwise
stated, the parameters are $\alpha=2$, $\gamma_{a}=0$, $\gamma_{p}/\pi=4.8\,$GHz,
$\eta/\pi=8.68\,$GHz, $\beta/\pi=6.75\,$GHz, $\tau_{f}=3.3\,$ns
and $\tau_{r}=3.5\,$ns. The X and Y components of the intensity read
\begin{eqnarray}
I_{x} & = & I_{0}\cos^{2}\left(\Phi/2\right)\quad,\quad I_{y}=I_{0}\sin^{2}\left(\Phi/2\right).
\end{eqnarray}

\begin{figure}
\begin{centering}
\includegraphics[bb=75bp 10bp 930bp 570bp,clip,width=1\columnwidth]{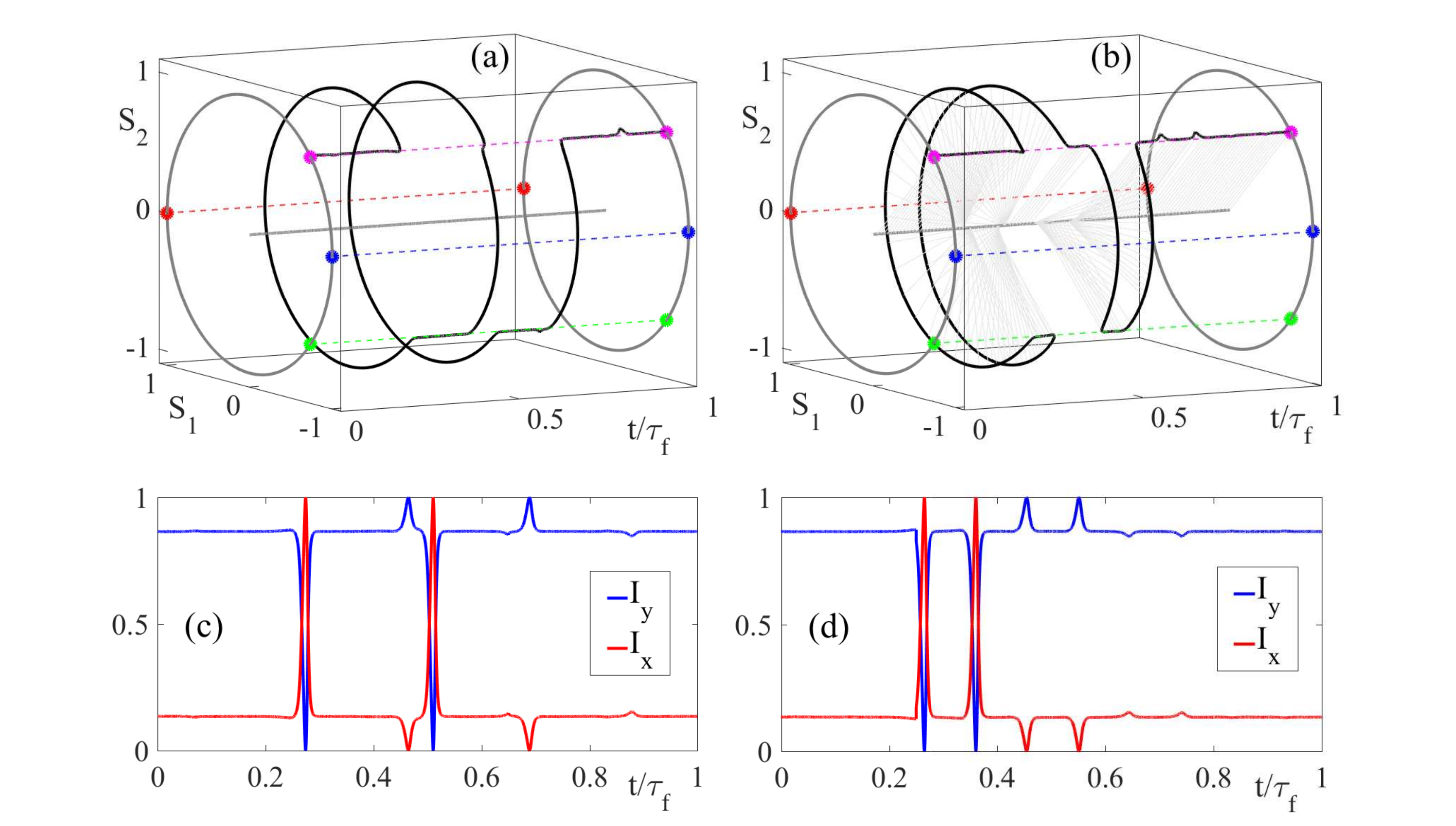}
\par\end{centering}
\centering{}\caption{Numerical integration of Eq.~(\ref{eq:SineDDE}). Evolution of the
Stokes parameters $\left(S_{1},S_{2}\right)$ and intensities $\left(I_{x},I_{y}\right)$
in the cases of a covalent (a,c) and catenane (b,d) molecule. Pure
emission along the X and Y directions correspond to the red and blue
lines with $\left(S_{1},S_{2}\right)=\left(1,0\right)$ and $\left(S_{1},S_{2}\right)=\left(-1,0\right)$,
respectively, while the steady state of Eq.~\ref{eq:SineDDE} $\Phi_{s}$
is depicted in pink while the green line represents $-\Phi_{s}$.\label{fig:KAK}}
\end{figure}

Figure~\ref{fig:setup}b,c) shows that, in each polarization, the
main pulse is followed by a small inverted kink after a time $\Delta\tau=\tau_{r}-\tau_{f}$.
This echo is the signature of the nonlocal coupling induced by the
additional delay $\tau_{r}$. It bears some similarity with the interaction
between temporal LSs in injected Kerr fibers mediated by sound-waves
\cite{JEM-NAP-13}, although here the effect is fully controllable
and, as it will be shown below, it can be used to tune the interaction
between LSs. These nonlocal echos create binding forces and allow
the existence of LS molecules whose separation between elements is
precisely $\Delta\tau$, as illustrated in Fig.~\ref{fig:sketch}b)
and demonstrated in \cite{MJB-NAP-15}. The structure of a typical
covalent molecule is depicted in terms of the Stokes parameters and
the polarization resolved intensities in Fig.~\ref{fig:KAK}a,c).
Each localized state is composed by a large polarization kink for
$\Phi$ followed by their nonlocal echo at $\Delta\tau$. The second
LS is linked to the first via its echo, such that the bonding distance
is $\Delta\tau$.

As illustrated in Fig.~\ref{fig:sketch}, when $\Delta\tau$ is sufficiently
large with respect to the pulse size, a new kind of molecule of interlaced
LSs can be observed where the second LS is trapped between the first
LS and its echo. This\emph{ }molecule of nested LSs is represented
in Fig.~\ref{fig:KAK}d) as a function of the polarization resolved
intensities and in Fig.~\ref{fig:KAK}b) in terms of the Stokes parameters.
After a large rotation corresponding to the primary peak of the first
LS, the system performs the opposed movement in correspondence with
the primary peak of the second LS, thus coming back to its original
state. Similar but smaller kinks follow as the nonlocal echoes of
the two LSs. Accordingly, Fig.~\ref{fig:KAK}b) evidences that this
nested molecule has the additional property of being composed of a
kink and an anti-kink, such that the total topological charge is zero.

\begin{figure}[t]
\begin{centering}
\includegraphics[bb=0bp 0bp 677bp 272bp,clip,width=1\columnwidth]{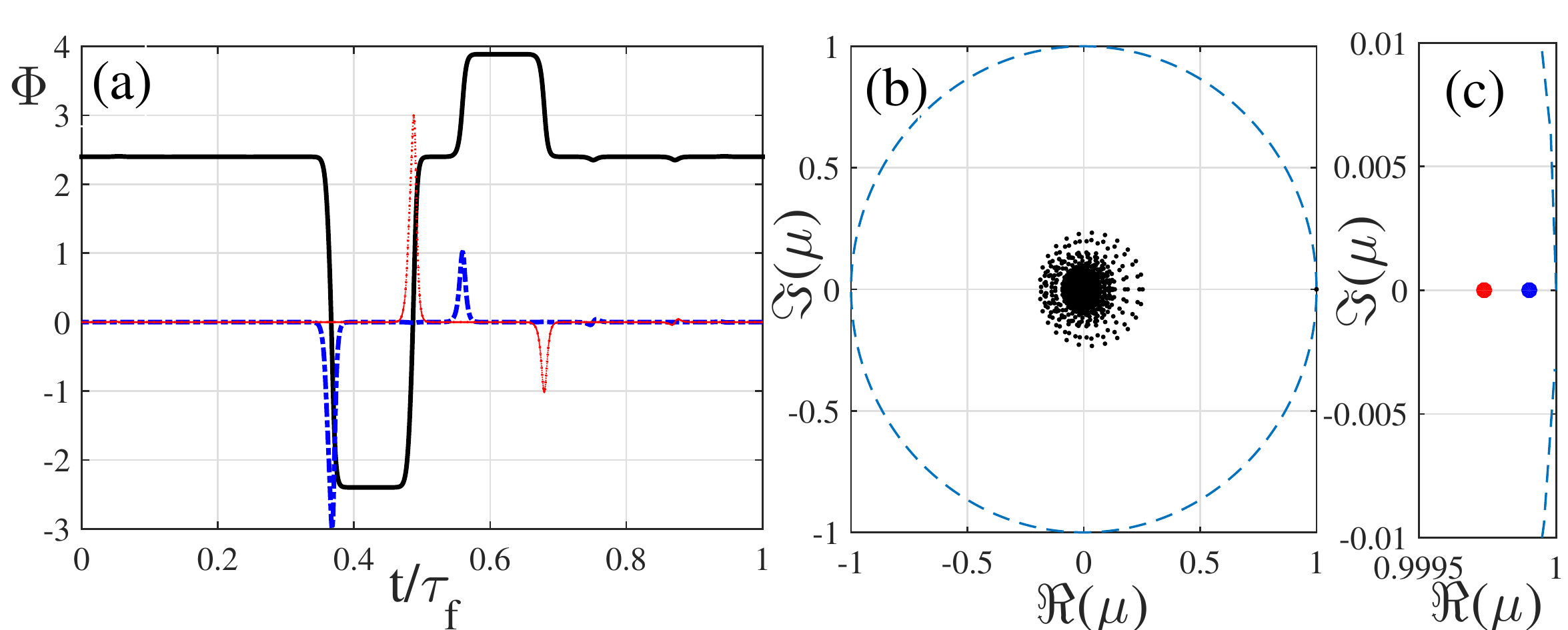}
\par\end{centering}
\begin{centering}
\includegraphics[bb=5bp 0bp 300bp 175bp,clip,width=1\columnwidth]{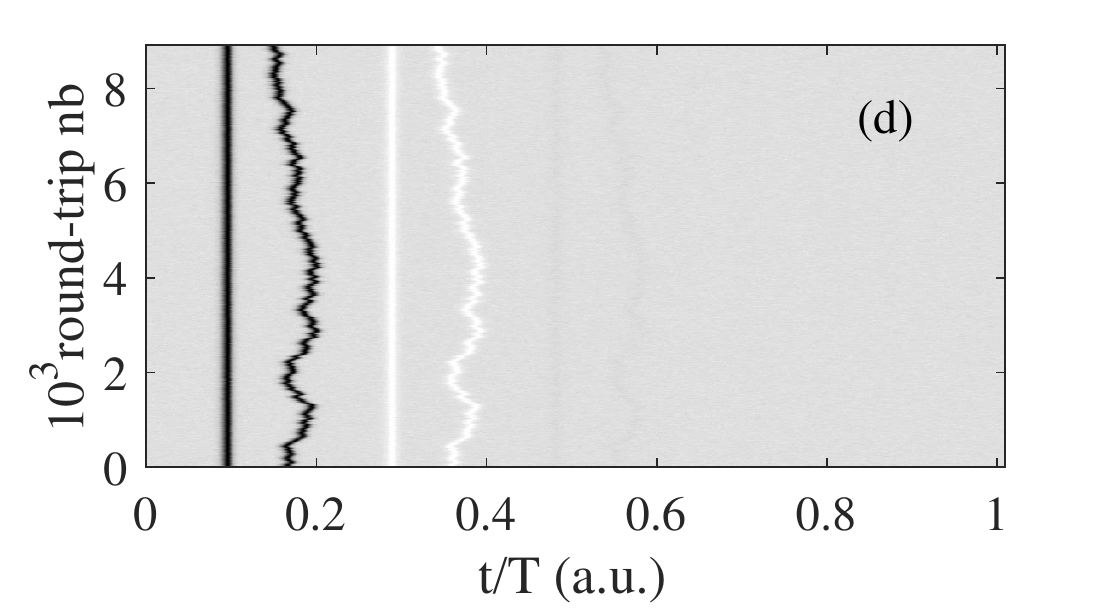}
\par\end{centering}
\centering{}\caption{Numerical analysis of the catenane molecule in Fig.~\ref{fig:KAK}b,d).
(a) Temporal profile (black) and the associated neutral eigenvectors
(dotted blue and dash-dotted red). (b) Floquet multipliers $\mu$
and (c) zoom around $\mu=1$ where one distinguishes two quasi-degenerate
multipliers. (d) Space-time diagram of \emph{\noun{$I_{y}$}} in presence
of additive white noise of amplitude $\xi=3\times10^{-2}$. The position
of each LS is given by the strong intensity dip followed by the smaller
intensity peak echoing at $\Delta\tau$. The motion of the two intensity
dips (dark peaks), mimicked by the corresponding echoes (white peaks),
is completely uncorrelated, thus evidencing the local independence
of the two LSs. The intensity of $I_{y}$ grows from black to white.
\label{fig:floquet_spacetime}}
\end{figure}

\begin{figure*}[!t]
\centering{}\includegraphics[bb=0bp 0bp 770bp
170bp,clip,width=2\columnwidth]{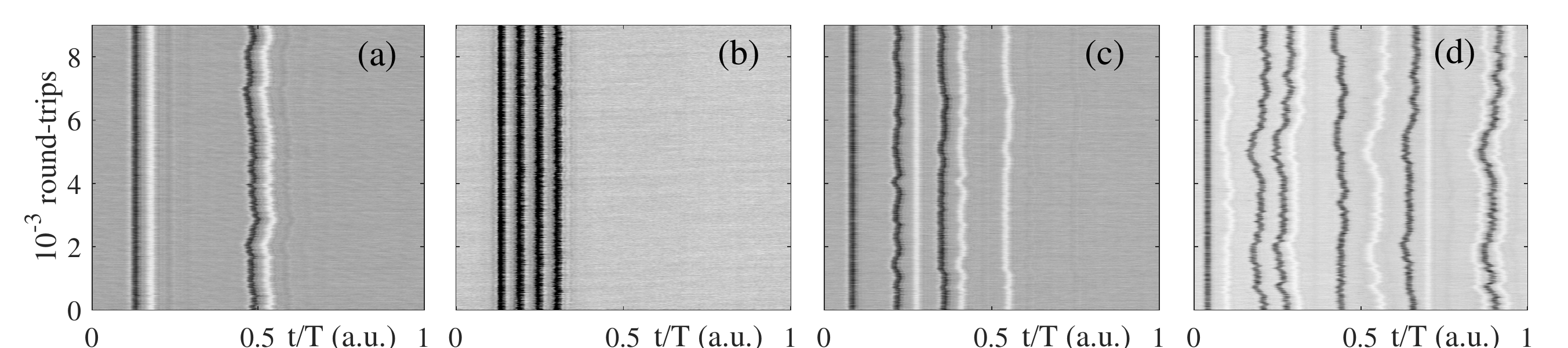}\caption{Experimental
space-time diagrams taking the first LS as a time reference for
visually enhancing the relative motion of other LSs. (a) two
independent LSs, $\tau_{r}=\tau_{f}+0.68\,$ns. (b) covalent molecule
with $\tau_{r}=\tau_{f}+0.33\,$ns and where the binding occurs via
the interaction with the nonlocal echo (drown inside the large
intensity dip). (b,c) Two different cases of catenane molecules with
(c) $\tau_{r}=2.3\,$ns and (d) $\tau_{r}=7.1\,$ns.
The intensity of $I_{y}$ grows from black to white. Common
parameters are $\tau_{f}=10.8\,$ns and $J=10J_{th}$.\label{fig:exp}}
\end{figure*}

Considering the LSs as periodic solutions of a high-dimensional dynamical
system permitted us to perform the analysis of their Floquet multipliers.
Floquet theory allows to study the linear stability of periodic solutions,
see for instance \cite{K-TE-08}. The stability analysis performed
in \cite{MJB-NAP-15} confirmed that the covalent molecule in Fig.~\ref{fig:KAK}a,c)
possesses a single neutral mode corresponding to the translation invariance
of the whole molecule. The results of a similar analysis applied to
the nested molecule in Fig.~\ref{fig:KAK}b,d) are summarized in
Fig.~\ref{fig:floquet_spacetime}. The temporal profile of the solution
over which we performed the stability analysis is depicted in Fig.~\ref{fig:floquet_spacetime}a)
and one notices in Fig.~\ref{fig:floquet_spacetime}b) not one but
\emph{two} quasi-degenerate Floquet multipliers close to $\mu=1$.
We also show in Fig.~\ref{fig:floquet_spacetime}a) the eigenvectors
associated to these \emph{two} neutral modes. One can easily identify
them with the temporal derivatives of the kink and of the anti-kink
composing the nested molecule, which allows for their individual translation.
The residual interactions between LSs, which are always present if
their separation is finite, renormalize the Floquet multipliers and
explain their small deviation with respect to unity. In the limit
$\left(\tau_{f},\tau_{r}\right)\rightarrow\left(\infty,\infty\right)$
we find numerically that they converge to $\mu=1$. In other words
our \emph{local} analysis shows that the two LSs composing the molecule
are indeed locally independent. However, it fails to show their global
dependence for which the consideration of the whole temporal profile
is needed. The local independence of the LSs forming the bound state
shown in Fig.~\ref{fig:KAK}b,d) contrasts with the rigid behavior
of the components of the molecule shown in Fig.~\ref{fig:KAK}a,c).
It also justifies our heuristic description of the molecule described
in Fig.~\ref{fig:sketch}c) as interlaced rings.

This evidence can be further supported by analyzing the motion of
the structure represented in Fig.~\ref{fig:KAK}b,d) over many
round-trips in presence of noise. As independent LSs and LS
molecules coexist for the same parameter values, they can only be
discriminated experimentally by observing their evolution on long
timescales. While independent LSs exhibit uncorrelated random walks
under the action of noise present in the system, LSs forming
covalent molecules behave as a unique rigid body, see for instance
Fig.~3a,b) in \cite{MJB-NAP-15}. The stochastic evolution of the
molecule described in Fig.~\ref{fig:KAK}b,d) is shown in
Fig.~\ref{fig:floquet_spacetime}d) by using a spatio-temporal
diagram where the time trace for the polarization resolved intensity
$I_{y}$ is folded over itself at each period $T\simeq\tau_{f}$.
Because we are plotting the intensity along the Y direction $I_{y}$,
the LS main pulse corresponds to a dip (D) while the echo
corresponds to small peak (P) at $\Delta\tau$ from D. In order to
help visualizing the fluctuations of the distance between the
molecule components, we have represented the LS evolution in the
reference frame where the first LS remains static. As a consequence,
while Fig.~\ref{fig:floquet_spacetime}d) does not provide any
information anymore regarding the random walk of the first LS, but
it magnifies the evolution of relative distance between the two LSs.
It also shows that, despite the fact that the two LSs can drift one
respect to the other, the distance between them remains bounded and
the main kink of the second LS is always caught between the main
kink of the first LS and its echo. Also, this representation allows
to distinguish between two elements catenanes and two independent
LSs. The signature of the first corresponds to two dips (D) followed
by two peaks (P), i.e. DDPP in Fig.~\ref{fig:floquet_spacetime}d),
while two independent LSs would correspond to DPDP.

Our theoretical predictions of the existence of catenane molecules
are supported by experimental observations. A wealth of these
molecule states has been observed in the experiment using the setup
described in Fig.~\ref{fig:setup}a), and they all coexist for the
same parameter values. Some examples are illustrated in
Fig.~\ref{fig:exp}. Besides two independent LSs in
Fig.~\ref{fig:exp}a), we represent in Fig.~\ref{fig:exp}b) a
standard covalent molecule found for small values of $\Delta\tau$,
where the binding occurs via the nonlocal echo, which corresponds to
the situation depicted in Fig.~\ref{fig:sketch}b). For larger values
of $\Delta\tau$, and besides the simplest catenane corresponding to
DDPP (not shown) we depict in Fig.~\ref{fig:exp}c) a situation where
three LSs are interlaced, giving the structures DDPDPP. Their local
independence can be deduced from their relative uncorrelated motion,
similar to that of independent LSs as depicted in
Fig.~\ref{fig:exp}a). Counting from left to right, we note that
LS$_{2}$ is interlaced with LS$_{1}$, while LS$_{3}$ is interlaced
with LS$_{2}$. A more complex catenane of 6 elements is shown in
Fig.~\ref{fig:exp}d). Here, two elements (LS$_{2}$ and
LS$_{3}$) are very close to each other and their echoes trap a distant
element (LS$_{6}$) driving its diffusion, thus evidencing the
binding forces induced by the echo at the source of catenane molecules.

In conclusion, we described how the presence of a pointwise nonlocality
in an extended system can give rise to a new kind of molecule of LSs
whose elements are simultaneously locally independent and globally
locked. Owing to the strong link between spatially extended and delayed
systems, we have analyzed the implementation of pointwise nonlocality
using a VCSEL enclosed in a double external cavity. The experimental
signature of the optical catenane is found to closely match our predictions.
We note that other TDSs capable of generating LSs as in \cite{MJB-PRL-14,GJT-NC-15},
would yield similar catenanes molecules.
\begin{acknowledgments}
J.J. acknowledges financial support project COMBINA (TEC2015-65212-C3-3-P
MINECO/FEDER UE) and the Ramón y Cajal fellowship. M.M and M.G. acknowledge
the Universitat de les Illes Baleares for funding a stay where part
of this work was developed.
\end{acknowledgments}


\end{document}